\documentclass[12pt]{article}
\usepackage{graphics}
\usepackage{epsfig}
\usepackage{latexsym}
\newcommand{\bfl}{{\bf l}}
\newcommand{\bfn}{{\bf n}}
\newcommand{\bfk}{{\bf k}}
\newcommand{\bfm}{{\bf m}}
\newcommand{\bfR}{{\bf R}}
\newcommand{\bfS}{{\bf S}}
\newcommand{\dpd}{\Delta_{pd}}          
\newcommand{\pd}{{\phantom{\dagger}}}
\newcommand{\cuotwo}{$C\!uO_2$}

\begin{document}
\thispagestyle{empty}
\begin{titlepage}
\centerline{\Large\bf Derivation of effective spin models}
\centerline{\Large\bf from a three band model for 
${\hbox{CuO}}_{\hbox{2}}$--planes
}
\vskip1cm
\centerline{E. M\"uller--Hartmann and A. Reischl}\smallskip
\centerline{Institute of Theoretical Physics, University of Cologne,}\par
\centerline{Z\"ulpicher Str 77, D-50937 Cologne, Germany}
\vskip1cm

{\bf Abstract.} The derivation of effective spin models describing the low 
energy magnetic properties of undoped \cuotwo--planes is reinvestigated. 
Our study aims at a {\it quantitative} determination of the parameters of
effective spin models from those of a multi--band model and is supposed to be
relevant to the analysis of recent improved experimental data on the spin
wave spectrum of $La_2CuO_4$. Starting from a conventional three--band model 
we determine the exchange couplings for the nearest and next--nearest neighbor 
Heisenberg exchange as well as for 4-- and 6--spin exchange terms via a direct 
perturbation expansion up to 12th (14th for the 4--spin term) order with 
respect to the copper--oxygen hopping $t_{pd}$. Our results demonstrate that 
this perturbation expansion does not converge for hopping parameters of the 
relevant size. Well behaved extrapolations of the couplings are derived, 
however, in terms of Pad\'e approximants. In order to check the significance 
of these results from the direct perturbation expansion we employ the 
Zhang--Rice reformulation of the three band model in terms of hybridizing 
oxygen Wannier orbitals centered at copper ion sites. In the Wannier notation 
the perturbation expansion is reorganized by an exact treatment of the strong 
site--diagonal hybridization. The perturbation expansion with respect to the 
weak intersite hybridizations is calculated up to 4th order for the 
Heisenberg coupling and up to 6th order for the 4--spin coupling. It shows 
excellent convergence and the results are in agreement with the Pad\'e 
approximants of the direct expansion. The relevance of the 4--spin coupling as 
the leading correction to the nearest neighbor Heisenberg model is emphasized.

\end{titlepage}

\section{Introduction}

After the discovery of the high--$T_c$ superconducting oxides \cite{bednorz}
it soon became clear that a minimum model for describing their electronic
properties had to contain at least three bands \cite{emery} derived from the
copper crystal field state $3d_{x^2-y^2}$ and from oxygen $2p_x$ and $2p_y$
orbitals \cite{andersen}. In a seminal paper Zhang and Rice showed 
\cite{zhang1} that the low energy physics of the three-band model is in fact 
contained in an effective single--band model, the type of model which was 
envisaged initially by Anderson \cite{anderson}.

The work presented here is concerned with a reinvestigation of the derivation
of effective single--band models from three-band models for \cuotwo--planes.
The present paper will be confined to the study of undoped \cuotwo--planes
where the effective models contain spin degrees of freedom only.
Effective low energy models are derived from high energy parent models via 
perturbative expansions \cite{takahashi}. The focus of this work is placed on 
how to obtain {\it high precision coupling constants} for the effective spin 
models and what are the leading corrections to the familiar nearest neighbor 
Heisenberg model. For the system of strongly correlated electrons considered 
here the copper-oxygen hopping $t_{pd}$ is the expansion parameter of choice. 
The expansion in powers of $t_{pd}$ is, however, not straightforward due to 
its rather small radius of convergence. Therefore, expansions beyond the 
leading order are required for obtaining reliable results. This is probably 
the reason why in existing derivations of the magnetic Hamiltonian of 
\cuotwo--planes the couplings are usually off by a factor of up to 2 
\cite{stein,yildirim}. 
The dominant term in the effective Hamiltonian is the Heisenberg 
nearest neighbor exchange obtained in fourth order which is substantially 
corrected by higher order contributions which we will present up to twelfth
order. In eighth order ring exchange processes start to contribute four--spin 
terms to the effective Hamiltonian which turn out to be not at all small 
\cite{schmidt}. 
Our results are consistent with the recent interpretation \cite{katanin} of 
improved experimental data on the spin wave spectrum of $La_2CuO_4$ in terms 
of sizable four--spin exchange terms \cite{coldea}. In comparison to these 
four--spin terms second and third neighbor Heisenberg terms which also first 
appear in eighth order turn out to be rather tiny. We have calculated all 
these terms up to twelfth order (four--spin term up to fourteenth order). It 
is evident from the results of these series expansions that physically 
relevant values of $t_{pd}$ are larger than the radius of convergence. We 
find, however, that Pad\'e approximants of these series expansions provide 
consistent extrapolations to the range of physically relevant model parameters.

There is an alternative approach to the perturbative treatment of three band 
models which shows a much better behavior of convergence and which we have also
applied to obtain an independent check of the significance of the Pad\'e
approximants derived from the direct expansion. This approach has been 
introduced in the paper of 
Zhang and Rice on the three band model in which this model was reformulated
in terms of hybridizing oxygen Wannier orbitals centered at the copper ion 
sites \cite{zhang1}. In this notation the hopping Hamiltonian contains a large 
site--diagonal hybridization $t_0$ which is easily treated exactly for each 
copper ion site and small intersite hybridizations which are then treated 
safely in a perturbative fashion. Along these lines Zhang and Rice achieved 
not only a clever rearrangement of the $t_{pd}$ perturbation series, but also a
particularly transparent formulation of the physics of doped \cuotwo--planes
in terms of ``spin'' and ``hole'' states the latter of which are known as 
Zhang--Rice singlets. In the effective low energy model (``$t$--$J$ model'') 
obtained this way neighboring ``spins'' experience an exchange interaction $J$ 
and ``holes'' interchange their position with neighboring ``spins'' via a 
hopping parameter $t$ \cite{dagotto}. We will show that the leading 
contribution to the nearest neighbor Heisenberg exchange obtained in second 
order in the intersite hopping is sufficient to reproduce the major features 
found from the direct expansion up to realistic values of $t_{pd}$, but is not 
sufficient for perfect agreement. Gided by sum rules for the hopping 
amplitudes in the Wannier representation we will then demonstrate how the 
agreement is systematically improved by including corrections of third and 
fourth order in the intersite hopping. The four--spin (up to sixth order) and 
further neighbor Heisenberg exchange terms will also be discussed in this 
context.

The paper is organized as follows. In the following section the three band 
model used in this work is briefly reviewed together with its transformation
into the Wannier representation. Section III describes the principles of the
perturbative derivation of effective Hamiltonians as we will use it. Section
IV is devoted to the direct expansion with respect to $t_{pd}$ and section V to
the expansion in the Wannier representation. The results are summarized and
conclusions are drawn in connection with the experimental evidence in section
VI.

\section{\bf The three--band model}

In this section we will briefly present the three--band model \cite{emery}
from which our investigation is going to start and fix the notations used. 
For the purpose of this paper which is focusing on the feasibility of high 
precision determination of the parameters of effective spin models we will 
use a minimum three-band model with the Hamiltonian 
\begin{equation}\label{hamil}
H=H_\epsilon+H_U+H_{pd}
\end{equation}
where the first term
\begin{equation}\label{heps}
H_\epsilon=\sum_{\bfl,\sigma}\bigl[\epsilon_dd^\dagger_{\bfl,\sigma}
d^\pd_{\bfl,\sigma}+\epsilon_p(p^\dagger_{x,\bfl+\bfn_x/2,\sigma}
p^\pd_{x,\bfl+\bfn_x/2,\sigma}+
p^\dagger_{y,\bfl+\bfn_y/2,\sigma}p^\pd_{y,\bfl+\bfn_y/2,\sigma})\bigr]
\end{equation}
describes the energies of the $3d$-- and $2p$--holes involved, the second
term
\begin{equation}\label{hu}
H_U=U\sum_{\bfl}d^\dagger_{\bfl,\uparrow}
d^\pd_{\bfl,\uparrow}d^\dagger_{\bfl,\downarrow}
d^\pd_{\bfl,\downarrow}
\end{equation}
describes the Coulomb repulsion of holes on the $Cu^{3+}$ ions and the third 
term
\begin{equation}\label{htpd}
H_{pd}=t_{pd}\sum_{\bfl,\sigma}\bigl[d^\dagger_{\bfl,\sigma}
(p^\pd_{x,\bfl+\bfn_x/2,\sigma}+p^\pd_{y,\bfl+\bfn_y/2,\sigma}
-p^\pd_{x,\bfl-\bfn_x/2,\sigma}-
p^\pd_{y,\bfl-\bfn_y/2,\sigma})+h.c.\bigr].
\end{equation}
describes the hopping of holes between $3d$-- and neighboring $2p$--sites. 
Copper $3d_{x^2-y^2}$--orbitals are placed on a square lattice in the 
$(x,y)$--plane which is spanned by unit vectors $\bfn_x$ and $\bfn_y$ and the 
vertices of which are labeled by the integer vector $\bfl$. Oxygen $2p_x$-- and
$2p_y$--orbitals are placed at the center of $x$-- and $y$--bonds, 
respectively, between neighboring lattice sites.

Typical parameters for the three-band model (\ref{hamil}) being used to model
\cuotwo--planes are \cite{mcmahan,hybertsen}
\begin{equation}\label{param}
\dpd\doteq\epsilon_p-\epsilon_d=3.6\,{\rm eV},\quad U=8\,{\rm eV},
\quad t_{pd}=1.3\,{\rm eV}.
\end{equation}
For a direct expansion with respect to the hopping parameter $t_{pd}$ the
Hamiltonian (\ref{hamil}) is decomposed into
\begin{equation}\label{decomp}
H=H_0^p+V^p\quad{\rm with}\quad H_0^p=H_\epsilon+H_U\quad{\rm and}
\quad V^p=H_{pd}.
\end{equation}
Although the hopping amplitude $t_{pd}$ is smaller than the charge transfer 
energy $\dpd$ and than the Coulomb energy $U$ it turns out that a
direct expansion of the parameters of an effective low energy model with 
respect to $t_{pd}$, i.e. an expansion in powers of $V^p$, does not work for 
the parameter set (\ref{param}). We will demonstrate this later explicitly and 
we will estimate the radius of convergence of such a direct expansion as 
$t_{pd}^c\approx U/16=0.5\,{\rm eV}$. We will therefore work out this expansion
to higher orders and will extract useful information from this expansion via
Pad\'e approximants.

Zhang and Rice \cite{zhang1} found an elegant way to reorganize the
perturbation expansion by reformulating the three--band model in terms of
hybridizing oxygen Wannier orbitals centered at the copper ion sites.
The reformulated model is obtained after transforming the hopping term into
momentum space re\-pre\-sentation using the Fourier transformed operators
\begin{equation}\label{fourd}
d^\dagger_{\bfl,\sigma}={1\over\sqrt{L}}\sum_{\bfk\in BZ}e^{-i\bfk\bfl}
d^\dagger_{\bfk,\sigma}
\end{equation}
and
\begin{equation}\label{fourp}
p^\dagger_{\alpha,\bfl+\bfn_\alpha/2,\sigma}=
{1\over\sqrt{L}}\sum_{\bfk\in BZ}e^{-i\bfk(\bfl+\bfn_\alpha/2)}
p^\dagger_{\alpha,\bfk,\sigma}\quad(\alpha=x,y),
\end{equation}
where $L$ denotes the number of unit cells. With the form factor
\begin{equation}\label{formf}
f(\bfk)\doteq2\sqrt{\sin^2{k_x\over2}+\sin^2{k_y\over2}}=
2\sqrt{1-{\cos k_x+\cos k_y\over2}}
\end{equation}
and the normalized hybridizing Wannier orbital in momentum space representation
\begin{equation}\label{wannier}
w^\pd_{\bfk,\sigma}\doteq2i
(\sin{k_x\over2}\cdot p^\pd_{x,\bfk,\sigma}+
\sin{k_y\over2}\cdot p^\pd_{y,\bfk,\sigma})/f(\bfk)
\end{equation}
the hopping term reads
\begin{equation}\label{htwk}
H_{pd}=t_{pd}\sum_{\bfk,\sigma}f(\bfk)(d^\dagger_{\bfk,\sigma}
w^\pd_{\bfk,\sigma}+w^\dagger_{\bfk,\sigma}d^\pd_{\bfk,\sigma}).
\end{equation}
Applying the Fourier transform (\ref{fourd}) to the Wannier operators
$w^\dagger_{\bfk,\sigma}$ mutually orthogonal real space Wannier orbitals 
$w^\dagger_{\bfl,\sigma}$ centered at the copper sites are obtained. In terms 
of these the hopping Hamiltonian finally takes the form
\begin{equation}\label{htwl}
H_{pd}=t_{pd}\sum_{\bfl,\bfm,\sigma}\bigl[T_{\bfl-\bfm}\,
d^\dagger_{\bfl,\sigma}w^\pd_{\bfm,\sigma}+h.c.\bigr],
\end{equation}
where the Fourier coefficients
\begin{equation}\label{tr}
T_\bfR\doteq{1\over L}\sum_\bfk f(\bfk)\,e^{i\bfk\bfR}=
\int_{BZ}{d^2\bfk\over(2\pi)^2}f(\bfk)\,e^{i\bfk\bfR}
\end{equation}
of the form factor (\ref{formf}) have the full symmetry of the square lattice.
Numerical values of these coefficients are given in Table 1.
\begin{center}
\begin{tabular}{|c|c|}\hline
\bfR & $T_\bfR$\\ \hline \hline
$(0,0)$ & 1.916183\\ \hline
$(\pm1,0)$,$(0,\pm1)$ & -0.280186\\ \hline
$(\pm1,\pm1)$ & -0.047013\\ \hline
$(\pm2,0)$,$(0,\pm2)$ & -0.027450\\ \hline
$(\pm2,\pm1)$,$(\pm1,\pm2)$ & -0.013703\\ \hline
\end{tabular}\\[0.5ex]
Table 1. Numerical values for $T_\bfR$
\end{center}
The coefficients $T_\bfR$ satisfy the sum rules
\begin{eqnarray}\label{sumrule}
s_\bfl\doteq\sum_\bfm T_\bfm T_{\bfl-\bfm}
&=&\langle f^2(\bfk)e^{i\bfk\bfl}\rangle_\bfk\nonumber\\
&=&\cases{4&($\bfl=(0,0)$)\cr-1&($\bfl=(\pm1,0)$ or $(0,\pm1)$)\cr0&(else).}
\end{eqnarray}
which we are going to use later. Obviously, the site--diagonal amplitude 
$T_{(0,0)}$ is much larger than all the other amplitudes and satisfies by 
itself the sum rule $s_{(0,0)}=4$ already to $91.8\%$. The amplitudes to the 4
first neighbors are almost 7 times smaller than $T_{(0,0)}$ and including them
the sum rule $s_{(0,0)}=4$ is missed by only $0.35\%$. The amplitudes to further
neighbors are much smaller again. One can show that in the limit of large
distances the amplitudes drop asymptotically like
\begin{equation}\label{asymp}
T_\bfR\sim\frac{-1}{2\pi R^3}\quad(R\to\infty).
\end{equation}
To write the Hamiltonian (\ref{heps}) also in terms of Wannier states
non--hybridizing $2p$--orbitals orthogonal to the Wannier orbitals $w$ have to
be introduced. In momentum space representation they are given by
\begin{equation}\label{nonhyb}
v^\pd_{\bfk,\sigma}\doteq2i
(\sin{k_y\over2}\cdot p^\pd_{x,\bfk,\sigma}-
\sin{k_x\over2}\cdot p^\pd_{y,\bfk,\sigma})/f(\bfk)
\end{equation}
and since the $2p$--basis sets $(p_x,p_y)$ and $(w,v)$ are unitarily
equivalent one obtains
\begin{equation}\label{hepsw}
H_\epsilon=
\sum_{\bfl,\sigma}\bigl[\epsilon_dd^\dagger_{\bfl,\sigma}
d^\pd_{\bfl,\sigma}+\epsilon_p(w^\dagger_{\bfl,\sigma}
w^\pd_{\bfl,\sigma}+v^\dagger_{\bfl,\sigma}
v^\pd_{\bfl,\sigma})\bigr].
\end{equation}
The Wannier representation in (\ref{htwl}) and (\ref{hepsw}) allows a
decomposition of the total Hamiltonian (\ref{hamil}) into 
\begin{equation}\label{decomw}
H=H_0^w+V^w
\end{equation}
where a major part of the
hopping term (\ref{htpd}) is incorporated in the unperturbed Hamiltonian.
Using the shorthand notation
\begin{equation}\label{t0}
t_0\doteq T_{(0,0)}t_{pd}\approx1.916\,t_{pd} 
\end{equation}
the unperturbed Hamiltonian is chosen as \cite{zhang1}
\begin{eqnarray}\label{h0}
H_0^w&=&\sum_{\bfl}h_\bfl\nonumber\\
h_\bfl&=&\sum_{\sigma}
\bigl[\epsilon_dd^\dagger_{\bfl,\sigma}d^\pd_{\bfl,\sigma}+
\epsilon_p w^\dagger_{\bfl,\sigma}w^\pd_{\bfl,\sigma}+
t_0(d^\dagger_{\bfl,\sigma}w^\pd_{\bfl,\sigma}+
w^\dagger_{\bfl,\sigma}d^\pd_{\bfl,\sigma})\bigr]\\
&&\quad
+Ud^\dagger_{\bfl,\uparrow}d^\pd_{\bfl,\uparrow}d^\dagger_{\bfl,\downarrow}
d^\pd_{\bfl,\downarrow}.\nonumber
\end{eqnarray}
The non--hybridizing orbital $v$ can be ignored altogether in the minimum
three--band model considered here since it is always completely filled. 
The local Hamiltonians $h_\bfl$ act independently at each site $\bfl$. They are
easily diagonalized exactly. The perturbative part of the total Hamiltonian 
is then given by the intersite hopping terms in Wannier representation
\begin{equation}\label{hv}
V^w=t_{pd}\sum_{\bfl,\bfm,\sigma}^{\bfl\ne\bfm}T_{\bfl-\bfm}\bigl[
d^\dagger_{\bfl,\sigma}w^\pd_{\bfm,\sigma}+
w^\dagger_{\bfl,\sigma}d^\pd_{\bfm,\sigma}\bigr]
\end{equation}
which are so small that they can safely be treated perturbatively for model
parameters as given by (\ref{param}). The separation of energy scales between 
$H_0^w$ and $V^w$ achieved through the use of the Wannier representation 
is so substantial that it is hard to understand why a controversy about the 
scenario proposed by Zhang and Rice \cite{zhang1} arose early on
\cite{reiter1,reiter2,zhang2,reiter3,pang} which was still quoted as 
unsettled in the review by Dagotto \cite{dagotto}. Leading order perturbative 
calculations using the Wannier representation were performed by many authors 
(see, e.g., \cite{zaanen,lovtsov,schuettler,jefferson,hayn,belinicher,feiner}).

\section{\bf Perturbative derivation of effective Hamiltonians}

The perturbative derivation of effective Hamiltonians for correlated electron 
systems has a long history the early stages of which were summarized by 
Takahashi in 1977 \cite{takahashi}. In this paper Takahashi 
presents a particularly transparent description of the method and gives an 
explicit solution for the effective Hamiltonian to arbitrary order. We will 
briefly recall Takahashi's approach here, because we are going to perform the
perturbation expansions in this paper using his formulation and because we wish
to avoid controversies about the proper use of the method like in 
\cite{oles,macdonald}.

It is assumed that the total Hamiltonian of a system is decomposed into
\begin{equation}\label{ham}
H=H_0+V.
\end{equation}
In the case of interest $H_0$ has a degenerate subspace $U_0$ of ground states
with energy $E_0$. On switching on the perturbation $V$ continuously the
subspace $U_0$ evolves continuously into the subspace $U$ of the corresponding
low energy eigenspace of $H$. Takahashi presents an explicit perturbative 
formula to all orders in $V$ for an isometric linear transformation $\Gamma$: 
$U_0\to U$ describing the mapping of $U_0$ onto $U$. In terms of $\Gamma$
the effective Hamiltonian is then given by
\begin{equation}\label{hameff}
H_{\rm eff}=\Gamma^\dagger H\Gamma.
\end{equation}
It acts in the subspace $U_0$ of unperturbed eigenstates of $H_0$ and has the
same spectrum as the perturbed Hamiltonian $H$. In view of the explicit 
perturbation series of $\Gamma$ it is a pure problem of book--keeping to set up
the perturbation series for $H_{\rm eff}$ to any required order. In terms of
the projection operator $P_0$ onto the ground state subspace $U_0$ and the
resolvent operator
\begin{equation}\label{resolv}
S\doteq-\frac{1-P_0}{H_0-E_0}
\end{equation}
the full perturbation expansion up to fourth order is given by
\begin{eqnarray}\label{perturb}
H_{\rm eff}&=&E_0P_0+P_0V\!P_0+P_0V\!SV\!P_0\nonumber\\
&&+P_0V\!SV\!SV\!P_0-\frac{1}{2}P_0V\!P_0V\!S^2V\!P_0
-\frac{1}{2}P_0V\!S^2V\!P_0V\!P_0\nonumber\\
&&+P_0V\!SV\!SV\!SV\!P_0-\frac{1}{2}P_0V\!S^2V\!P_0V\!SV\!P_0
-\frac{1}{2}P_0V\!SV\!P_0V\!S^2V\!P_0\nonumber\\
&&+\frac{1}{2}P_0V\!P_0V\!P_0V\!S^3V\!P_0
+\frac{1}{2}P_0V\!S^3V\!P_0V\!P_0V\!P_0\\
&&-\frac{1}{2}P_0V\!P_0V\!S^2V\!SV\!P_0
-\frac{1}{2}P_0V\!SV\!S^2V\!P_0V\!P_0\nonumber\\
&&-\frac{1}{2}P_0V\!P_0V\!SV\!S^2V\!P_0
-\frac{1}{2}P_0V\!S^2V\!SV\!P_0V\!P_0.\nonumber
\end{eqnarray}
For the purposes of the calculations in this paper we had to list this 
expansion up to twelfth order. The number of terms in the series grows
exponentially with the order. To twelfth order the perturbation series 
contains 363721 terms.

For useful applications of the formal series of $H_{\rm eff}$ the unperturbed
Hamiltonian $H_0$ has to be easily diagonalized such that matrix elements of
the resolvent (\ref{resolv}) can be calculated explicitly. In this paper we 
will apply the perturbation expansion to the two Hamiltonian decompositions
(\ref{decomp}) and (\ref{decomw}) were this condition on $H_0$ is satisfied. 
We also will confine the analysis to undoped systems which implies that all
terms in $H_{\rm eff}$ containing $P_0VP_0$ don't contribute. This reduces
the number of twelfth order terms in $H_{\rm eff}$ to 12341. In the direct 
expansion based on (\ref{decomp}) ground states can only be connected by an 
even number of hopping processes such that all terms with any odd number of 
$V$ between two $P_0$ don't contribute. This reduces the number of twelfth
order terms to 3180. For the Wannier decomposition (\ref{decomw}) our 
analy\-sis will be confined to sixth order. In this case, of the terms given 
in (\ref{perturb}) only the second order term, the first third order term and 
the first three fourth order terms will contribute and up to sixth order 30 
terms have to be taken into account. Notice that in the Wannier
decomposition (\ref{decomw}) odd order terms do contribute since $H_0^w$ mixes 
$d$-- and $w$--orbitals.

\section{\bf Direct perturbation expansion}

In this section we are going to discuss the direct expansion with respect to
$t_{pd}$ on the basis of the decomposition (\ref{decomp}). In the undoped
case that we are considering here the subspace of ground states $U_0$ of the 
unperturbed Hamiltonian $H_0^p$ contains all states without any $p$--holes and 
with a single $d$--hole on each copper site. The effective Hamiltonian acting 
on $U_0$ is thus a pure spin Hamiltonian acting on the spins $S=1/2$ of the 
$d$--sites. Due to the symmetry properties of the three band model 
(\ref{hamil}) this Hamiltonian has got to be invariant under global spin 
rotations and under the space group of the square lattice. Only terms with an
even number of spins are possible due to time reversal symmetry. The excited 
states of $H_0^p$ are very simple and the excitation energies contain a 
Coulomb energy $U$ for each $d$--site with two holes and a charge transfer 
energy $\dpd$ for each $p$--hole. For a contribution of order n to the 
effective Hamiltonian one has to consider all sets of n hopping processes 
each of which defines a certain cluster of sites involved. Due to the linked 
cluster theorem (which is bound to hold to keep the effective Hamiltonian 
extensive) only connected clusters are known to contribute. It is therefore 
sufficient to evaluate the various orders of $H_{\rm eff}$ on certain finite 
clusters. We have implemented the purely symbolic evaluation of the series
expansion with a $C\!\!+\!\!+$ program.

For simplicity we will disregard any constant energy shift in $H_{\rm eff}$ 
since we want to focus on the effective spin Hamiltonian. The leading term in 
$H_{\rm eff}$ is then a fourth order nearest neighbor Heisenberg exchange 
$J_1\,\bfS_1\cdot\bfS_2$ with the well known exchange coupling (see e.g. 
\cite{zaanen})
\begin{equation}\label{jdir}
J^{(4)}_{1,\rm dir}
=\frac{2\,t_{pd}^4}{\dpd^2}\bigl(\frac{4}{2 \dpd} + \frac{2}{U}\bigr)
=\frac{4\,t_{pd}^4\,\left( U + \dpd \right) }{U\,\dpd^3}.
\end{equation}
In order to determine this coupling it is sufficient to calculate the amplitude
of a spin flip process on a three--site cluster consisting of two neighboring
$d$--sites and the $p$--site in between. In view of the identity
$\bfS_1\cdot\bfS_2=S_1^zS_2^z+\frac{1}{2}(S_1^+S_2^-+S_1^-S_2^+)$ the coupling 
is given by twice the spin flip amplitude.

In sixth order processes the six additional $p$--sites adjacent to the two 
$d$--sites can be visited by a hole. Therefore a nine site cluster would be 
sufficient to calculate $J^{(6)}_{1,\rm dir}$. Since in each individual 
exchange process at most one of the additional $p$--sites is visited the 
actual calculation can be confined to clusters of up to no more than four 
sites. Each of the six four--site clusters gives the same contribution to the 
sixth order coupling. In one such contribution either all four sites or only 
the three sites of the fourth order cluster will be involved in the exchange 
process. Therefore, the sixth order spin flip amplitude is given by six times 
the spin flip amplitude of the four--site cluster minus 5 times the spin flip 
amplitude of the three--site cluster. This type of reasoning would be 
dispensable in the sixth order case for which it was examplified here, but it 
is absolutely essential to make the higher order calculations feasible. It 
allows to reduce the maximum cluster size for the calculation of the nearest 
neighbor exchange from 17 to 8 in eighth order, from 31 to 9 in tenth order 
and from 43 to 12 in twelfth order.

In eighth order ring exchange processes on an eight--site plaquette visiting 
four $d$--sites are possible. These processes produce four--spin exchange terms
in $H_{\rm eff}$. In cases where multi--spin terms are present the fewer--spin 
exchange terms can be inferred in the following way. Partial traces (i.e. 
traces over some of the spins) of any multi--spin term vanish. By forming the
trace over some of the spins belonging to a cluster all exchange terms 
containing these spins are therefore projected out. Applying this reasoning
to the eight--site plaquette one obtains the two--spin exchange of a pair of
spins by averaging over all configurations of the other spins contained in the 
plaquette. From time reversal invariance and hermiticity of $H_{\rm eff}$ one
can infer that the amplitude of a spin flip process remains unchanged if all
unflipped spins of a cluster are inverted. This allows to reduce by a factor of
2 the number of configurations needed for the averaging.

Along the lines described above we have calculated the nearest neighbor 
exchange coupling $J_{1,{\rm dir}}^{\vphantom{4}}$ up to twelfth order in 
$t_{pd}$. The full formula of the twelfth order result is given by 
Eq.~(\ref{J1Ordn10}) in Appendix A. Fig.~1 shows how the ratio 
$J_{1,{\rm dir}}^{\vphantom{4}}/J_{1,{\rm dir}}^{(4)}$ varies with increasing 
$t_{pd}$ if the sixth, eighth, tenth and twelfth order terms are 
included (see the thick lines in Fig.~1). It is 
\begin{figure}[h]
\begin{center}
\epsfxsize=12cm
\epsfbox{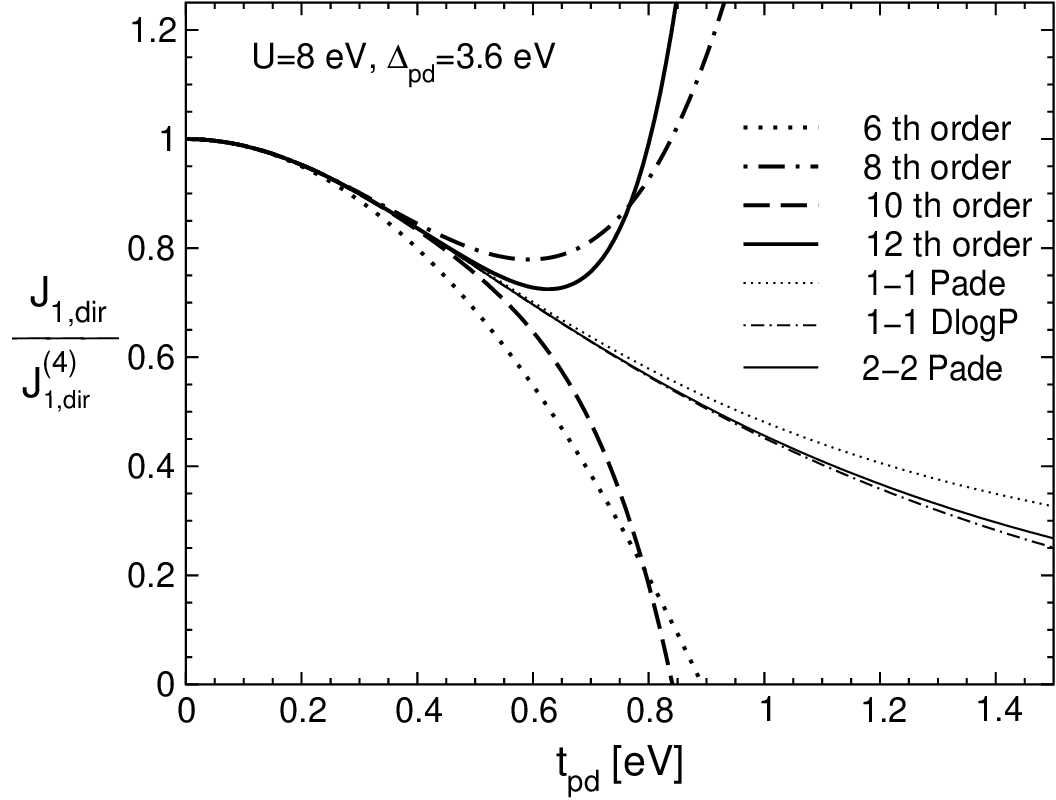}
\end{center}
\caption{Variation of $J_{1,{\rm dir}}/J_{1,{\rm dir}}^{(4)}$ with $t_{pd}$.}
\label{fig1}
\end{figure}
obvious from this plot that 
for the physically relevant values of $t_{pd}$ as given in (\ref{param}) $J_1$ 
is smaller than simple estimates from $J_{1,{\rm dir}}^{(4)}$ would suggest, 
but it is also obvious that the radius of convergence of the direct 
perturbation series considered here is much smaller than 
$t_{pd}=1.3\,{\rm eV}$. The direct series determines $J_1$ accurately only up
to $t_{pd}\approx0.5\,{\rm eV}$. Extrapolations beyond the radius of 
convergence can be obtained, however, via Pad\'e approximants of the series 
for $J_{1,{\rm dir}}^{\vphantom{4}}/J_{1,{\rm dir}}^{(4)}$. We denote as 
m-n~Pad\'e the approximant with an ${\rm m^{th}}$ order numerator polynomial 
and an ${\rm n^{th}}$ order denominator polynomial in the variable 
$x=t_{pd}^2$. We have also constructed extrapolations via analogous Pad\'e 
approximants for the logarithmic derivative of
$J_{1,{\rm dir}}^{\vphantom{4}}/J_{1,{\rm dir}}^{(4)}$ which we denote by
m-n~DlogPad\'e. The 1-1 and 2-2~Pad\'es and the 1-1~DlogPad\'e shown by the 
thin lines in Fig.~1 demonstrate the excellent convergence of this 
extrapolation procedure. The 0-1, 1-3 and 1-2~Pad\'es and the 1-2 and 
2-1~DlogPad\'es are not shown because they all differ from the 2-2~Pad\'e by 
less than $3\%$ for $t_{pd}\le1.3\,{\rm eV}$ and less than $4\%$ for 
$t_{pd}\le1.5\,{\rm eV}$. From this observation we derive the estimate that 
they determine the nearest neighbor exchange coupling with an accuracy of 
better than $4\%$. Note the substantial reduction of the coupling in the 
range of physical interest, $J_1=0.33\,J_{1,{\rm dir}}^{(4)}$ for 
$t_{pd}=1.3\,{\rm eV}$, in comparison to the lowest order result. 

The four--spin exchange terms which first appear in eighth order can be 
inferred from considering processes in which all four spins are flipped. Let us
label the spins on the four $d$--sites of a square plaquette in cyclic order by
numbers 1 to 4. There are 3 independent four--spin invariants, 
$(\bfS_1\cdot\bfS_2)(\bfS_3\cdot\bfS_4)$, 
$(\bfS_2\cdot\bfS_3)(\bfS_4\cdot\bfS_1)$ and
$(\bfS_1\cdot\bfS_3)(\bfS_2\cdot\bfS_4)$ from which the four--spin exchange 
terms have to be formed. Due to the square point symmetry of our model the 
first two invariants always get the same exchange coupling in the 
effective Hamiltonian. This common coupling is given by twice the
amplitude of the process which flips all spins of the initial state
$|1\!\uparrow,2\!\downarrow,3\!\uparrow,4\!\downarrow\rangle$, since the third 
invariant doesn't contribute to this process. The exchange coupling of the
third invariant can be inferred from considering an alternative four--spin flip
process starting from the initial state 
$|1\!\uparrow,2\!\downarrow,3\!\downarrow,4\!\uparrow\rangle$. The sum
of the couplings of the first invariant and the third invariant is given by
four times the amplitude of this process. It turns out that this amplitude
vanishes in eighth and tenth order. This implies that up to tenth order the
four--spin exchange term has the form
\begin{equation}\label{squareex}
J_{\Box}\,\Big[(\bfS_1\cdot\bfS_2)(\bfS_3\cdot\bfS_4)+
(\bfS_2\cdot\bfS_3)(\bfS_4\cdot\bfS_1)-
(\bfS_1\cdot\bfS_3)(\bfS_2\cdot\bfS_4)\Big]
\end{equation}
in analogy to what is known for the one band Hubbard model in fourth order
\cite{takahashi}.

The vanishing of the 
$|1\!\uparrow,2\!\downarrow,3\!\downarrow,4\!\uparrow\rangle$ spin flip process
up to tenth order can be easily understood as resulting from the linked 
cluster theorem, because for these processes the plaquette (1,2,3,4) decomposes
into two unlinked clusters, one of them containing the $d$--sites 1 and 2, the 
other containing sites 3 and 4. In twelfth order there are processes linking 
these two clusters and producing another four--spin term 
\begin{equation}\label{timesex}
J_{\times}\,(\bfS_1\cdot\bfS_3)(\bfS_2\cdot\bfS_4)
\end{equation}
in addition to (\ref{squareex}). It has to be noted that in twelfth order 
clusters containing 6 $d$--sites are created which produce six--spin terms in 
the effective Hamiltonian. In the calculation of the four--spin terms these 
six--spin terms have to be properly eliminated by the averaging procedure 
described above. 

The eighth order coupling constant of the four--spin term (\ref{squareex}) is 
found to be
\begin{equation}\label{jsquare}
J_{\Box,{\rm dir}}^{(8)}=\frac
{80\,t_{pd}^8\,( U + \dpd) \,( U^2 + U\,\dpd + \dpd^2)}{U^3\,\dpd^7}.
\end{equation}
Corrections up to fourteenth order are shown by Eq.~(\ref{JsqOrdn12}) in 
Appendix A together with the leading order contribution for $J_\times$ in 
Eq.~(\ref{JxOrdn12}). The variation of 
$J_{\Box,{\rm dir}}^{\vphantom{4}}/J_{\Box,{\rm dir}}^{(8)}$ with increasing
$t_{pd}$ is shown in Fig.~2. Here, the 0-3~Pad\'e (not shown) and the 1-2 and 
the 2-1~Pad\'es as well as the 1-1~DlogPad\'e seem to provide a rather accurate
estimate with an uncertainty of about $\pm6\%$ for $t_{pd}=1.3\,{\rm eV}$ and
an uncertainty of about $\pm15\%$ for $t_{pd}=1.5\,{\rm eV}$. For 
$t_{pd}=1.3\,{\rm eV}$ the coupling $J_\Box$ is about 10 times smaller than 
suggested by the leading order term. We will consider the 1-1~DlogPad\'e the 
most probable estimate of
$J_{\Box,{\rm dir}}^{\vphantom{4}}/J_{\Box,{\rm dir}}^{(8)}$.
\begin{figure}[h]
\begin{center}
\epsfxsize=12cm
\epsfbox{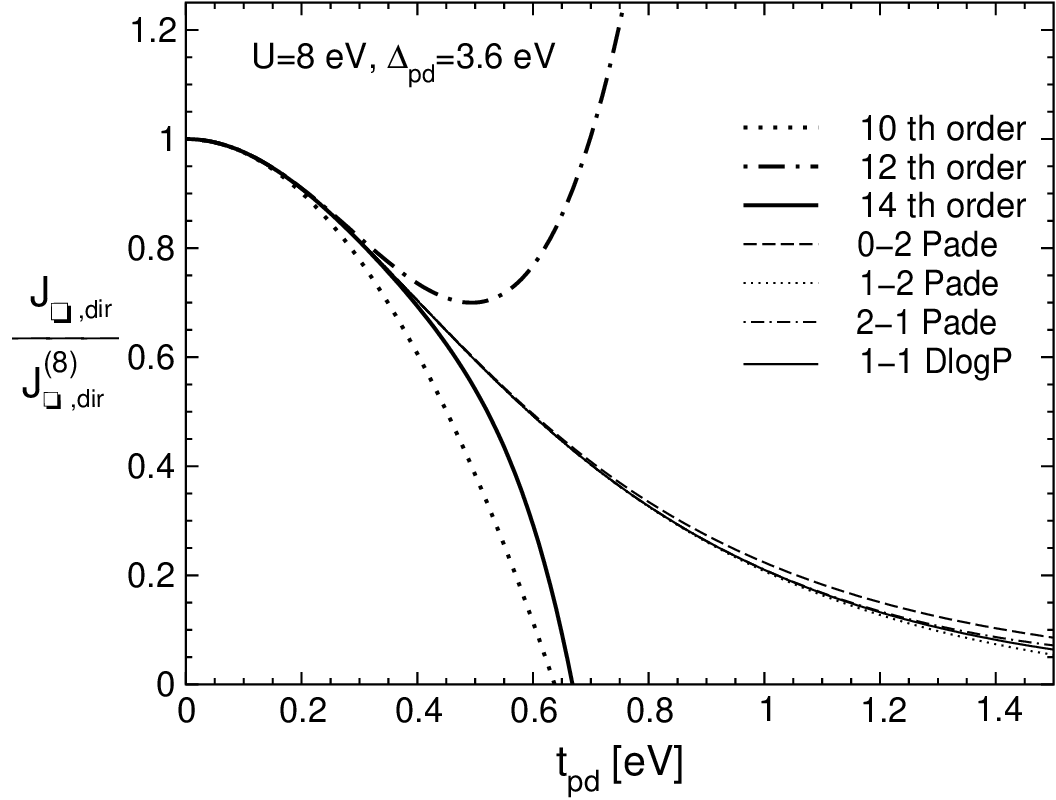}
\end{center}
\caption{Variation of $J_{\Box,{\rm dir}}/J_{\Box,{\rm dir}}^{(8)}$ with 
$t_{pd}$.}
\label{fig2}
\end{figure}

The leading contributions to second neighbor Heisenberg exchange terms like 
$J_2\,\bfS_{(0,0)}\cdot\bfS_{(1,1)}$ and to third neighbor Heisenberg exchange 
terms like $J_3\,\bfS_{(0,0)}\cdot\bfS_{(0,2)}$ are also obtained in eighth 
order. These couplings are given by
\begin{equation}\label{jtwo}
J_{2,{\rm dir}}^{(8)}=\frac{4\,t_{pd}^8\,
\left( 11\,U^3 + 4\,U^2\,\dpd + 2\,U\,\dpd^2 + \dpd^3 \right)}
{U^3\,\dpd^7}
\end{equation}
and
\begin{equation}\label{jthree}
J_{3,{\rm dir}}^{(8)}=\frac{4\,t_{pd}^8\,\left( 3\,U^3 + 2\,U^2\,\dpd + 2\,U\,
\dpd^2 + \dpd^3 \right) }{U^3\,\dpd^7},
\end{equation}
respectively. Corrections to these leading order expressions which we have
calculated to twelfth order are given by Eqs. (\ref{J2Ordn12}) and 
(\ref{J3Ordn12}) in Appendix A. 

\begin{figure}[h]
\begin{center}
\epsfxsize=12cm
\epsfbox{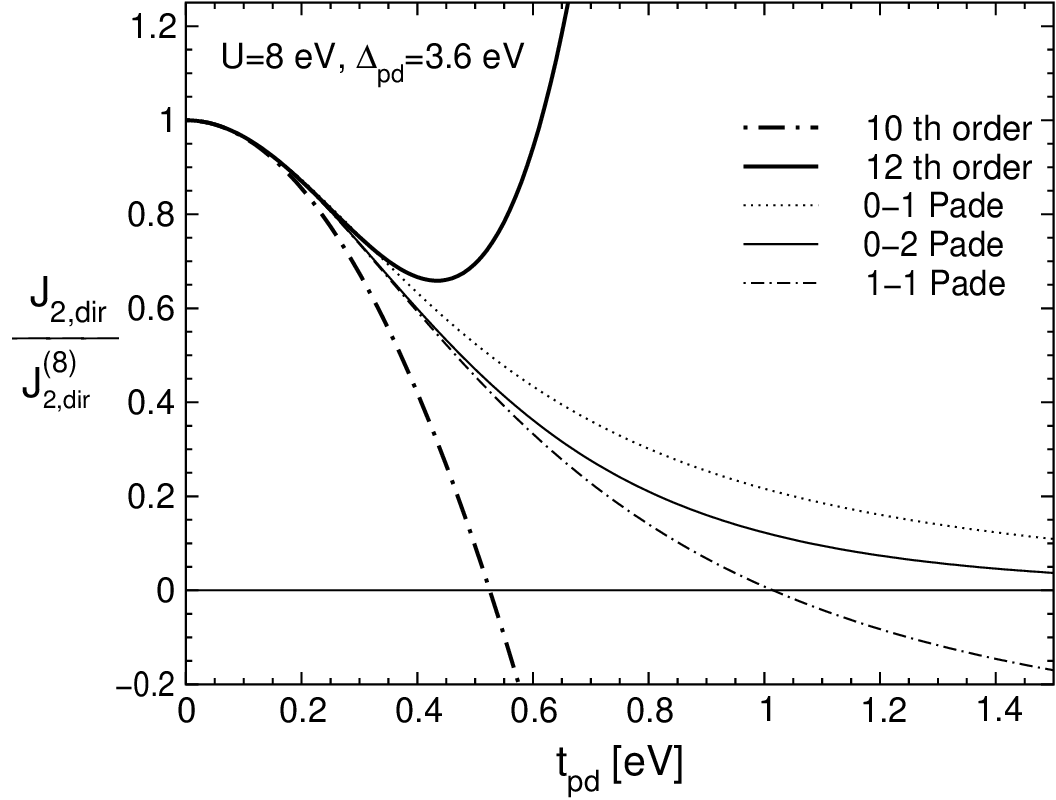}
\end{center}
\caption{Variation of $J_{2,{\rm dir}}/J_{2,{\rm dir}}^{(8)}$ with $t_{pd}$.}
\label{fig3}
\end{figure}

Figs.~3 and 4 show the variation of 
$J_{2,{\rm dir}}^{\vphantom{4}}/J_{2,{\rm dir}}^{(8)}$ and
$J_{3,{\rm dir}}^{\vphantom{4}}/J_{3,{\rm dir}}^{(8)}$, respectively, with
increasing $t_{pd}$. The radii of convergence appear to be even smaller than
in the case of $J_1$. The 1-1~Pad\'e in Fig.~3 might indicate that $J_2$ 
changes sign slightly below $t_{pd}=1\,{\rm eV}$, but the scattering of the
various approximants doesn't allow definite conclusions on a change of sign.
Since the 0-1~DlogPad\'e (not shown in Fig.~3) coincides to high precision with
the 0-2~Pad\'e we will consider this approximant as the most probable estimate
for $J_2$. The 0-2~Pad\'e for $J_3$ shown in Fig.~4 turns upwards and has a 
pole at $t_{pd}\approx2.1{\rm eV}$. The other three approximants shown appear 
to behave consistently and we will consider the 0-1~DlogPad\'e as the most 
probable estimate for $J_3$. Altogether, the Pad\'e approximants for $J_2$ and 
$J_3$ scatter much more than those for $J_1$ and $J_\Box$ and provide less 
accurate estimates for $J_2$ and $J_3$. We do, however, learn from these 
extrapolations that for $t_{pd}=1.3\,{\rm eV}$ both $J_2$ and $J_3$ also are 
reduced substantially in comparison to the leading order results (\ref{jtwo}) 
and (\ref{jthree}), $J_2$ probably by a factor of as much as 10 and $J_3$
probably by a factor of 5. As we will see later $J_2$ and $J_3$ are so small in
absolute size that their accurate determination is less urgent for practical 
purposes.
\begin{figure}[h]
\begin{center}
\epsfxsize=12cm
\epsfbox{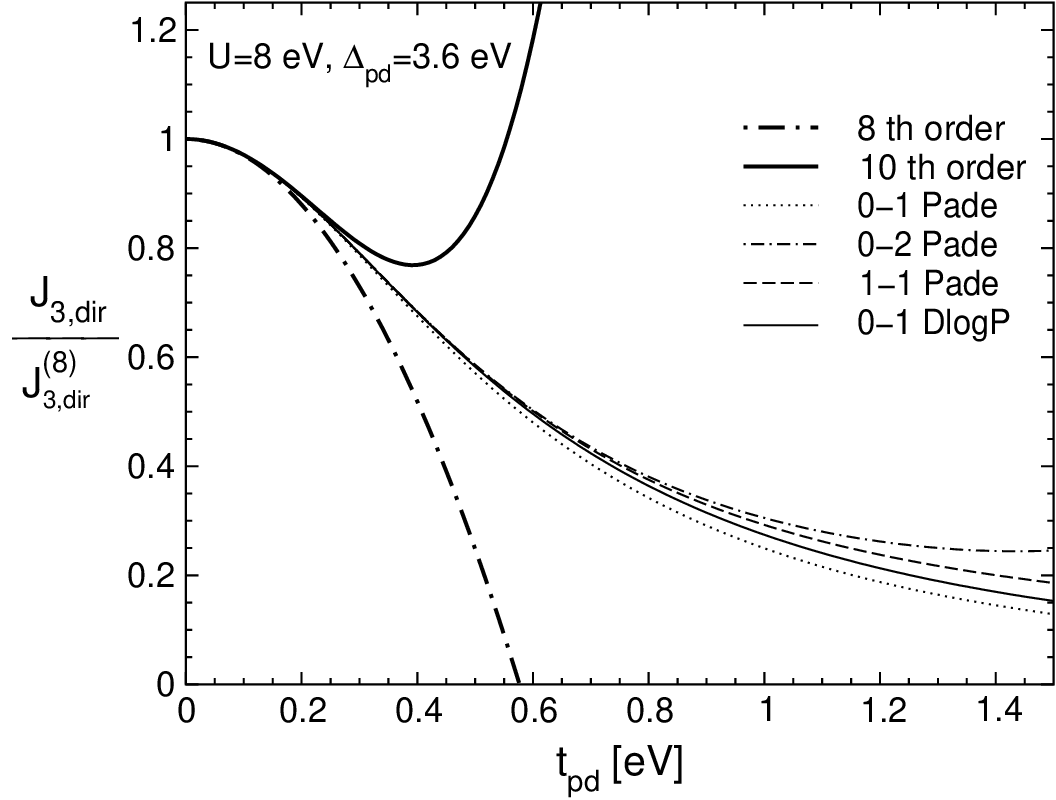}
\end{center}
\caption{Variation of $J_{3,{\rm dir}}/J_{3,{\rm dir}}^{(8)}$ with $t_{pd}$.}
\label{fig4}
\end{figure}

To describe the six--spin term resulting from twelfth order ring exchange 
processes on a double plaquette we label the six spins involved cyclically by
numbers 1 to 6. Since the oxygen ion at the center of the double plaquette is
not visited in twelfth order the six--spin term has the full symmetry of the
hexagon formed by the six spins. The 15 independent invariants obtained by all
pairings of the six spins into three scalar products \cite{spat} group into
the 5 operators with hexagonal symmetry $O_1$ to $O_5$ given by 
Eq.~(\ref{opsix}) in Appendix A. 
With the same type of arguments which led to (\ref{squareex}) we conclude that
the twelfth (and fourteenth) order six--spin term has the form
\begin{equation}\label{sixex}
J_{\sqsubset\!\sqsupset}(O_1+O_2-O_3+O_4-O_5).
\end{equation}
The exchange coupling $J_{\sqsubset\!\sqsupset}^{(12)}$ given by
Eq.~(\ref{Jsix}) in Appendix A was calculated from ring exchange processes 
which flip all spins of the state 
$|1\!\uparrow,2\!\downarrow,3\!\uparrow,4\!\downarrow,5\!\uparrow,
6\!\downarrow\rangle$.
\begin{figure}[h]
\begin{center}
\epsfxsize=12cm
\epsfbox{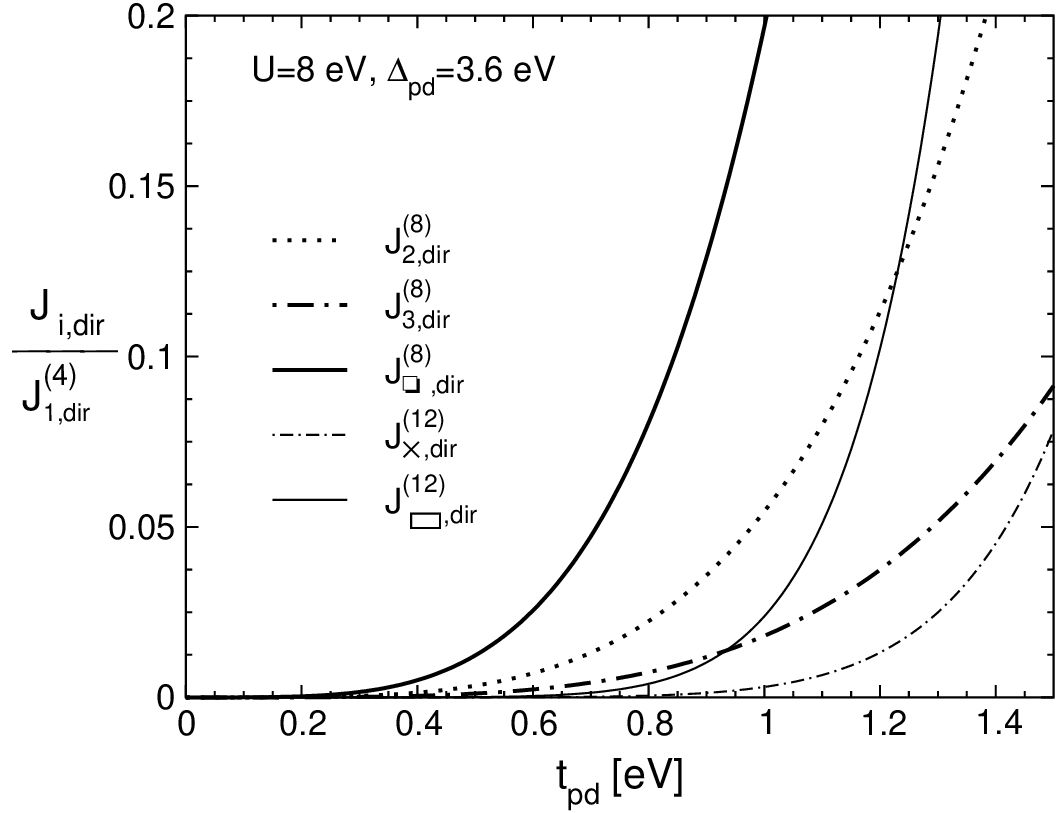}
\end{center}
\caption{Comparison of the leading order terms of the various couplings.}
\label{fig5}
\end{figure}

For a comparison of the relative sizes of the various couplings we first show
in Fig.~5 the leading perturbative contributions of all couplings determined,
in units of $J_{1,{\rm dir}}^{(4)}$. In the range of physically relevant model
parameters the four--spin coupling $J_\Box$ is by far the largest correction to
the nearest neighbor two--spin coupling $J_1$. The second and third neighbor
Heisenberg couplings $J_2$ and $J_3$ are much smaller and are in fact
comparable to the six--spin coupling $J_{\sqsubset\!\sqsupset}$. This scenario 
agrees with what is known from perturbation expansions for the single band 
Hubbard model \cite{takahashi,macdonald} and from cluster calculations for the 
three band model \cite{schmidt}. 

A quantitative comparison of the best approximants for the various 
couplings with $J_1$ (represented by its 2-2~Pad\'e) is shown in Fig.~6 
where we have denoted the m-n~Pad\'e for the coupling $J_i$ by $J_i\,[m,n]$. 
For the model parameters (\ref{param}), $J_\Box$ is almost one order of
magnitude smaller than $J_1$ and the couplings $J_2$ and $J_3$ are almost
another order of magnitude smaller. The four--spin coupling $J_\Box$ therefore
has to be considered an important modification of the simple nearest neighbor
Heisenberg model, whereas the second and third neighbor Heisenberg couplings
$J_2$ and $J_3$ (as well as the six--spin coupling $J_{\sqsubset\!\sqsupset}$) 
may be ignored as correction at the level of about $3\%$.
\begin{figure}[h]
\begin{center}
\epsfxsize=12cm
\epsfbox{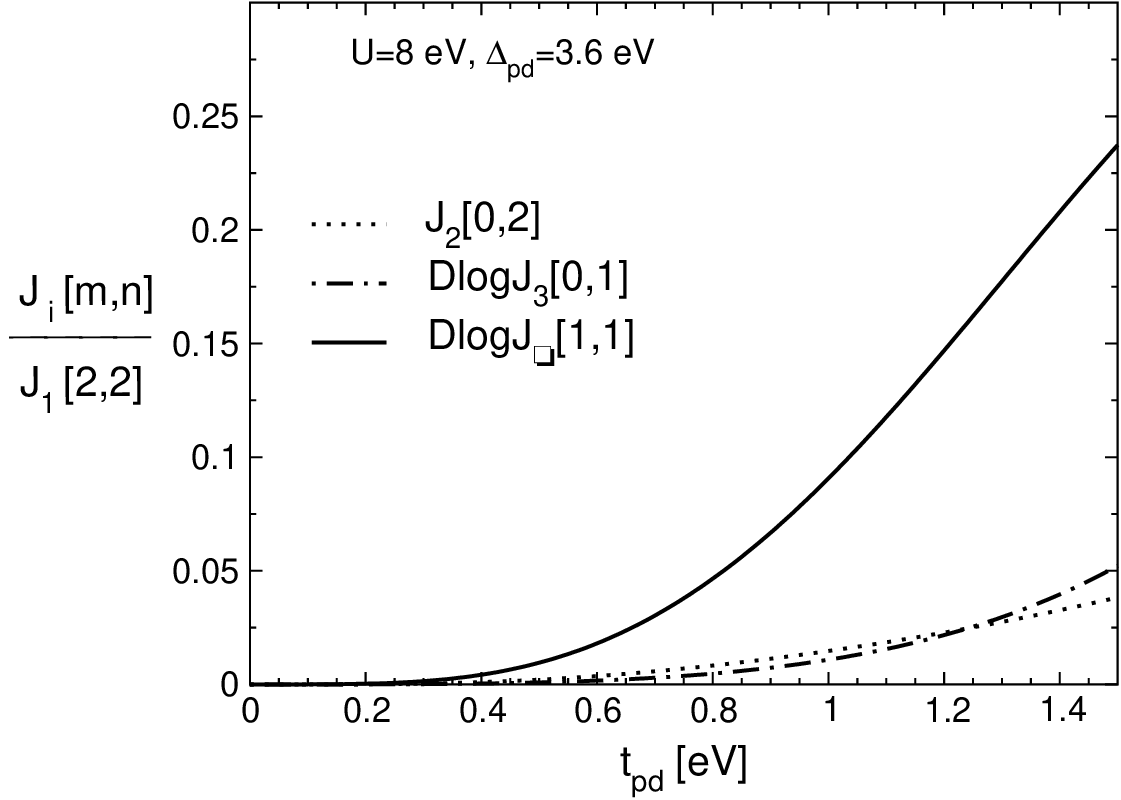}
\end{center}
\caption{Comparison of the best Pad\'e approximants.}
\label{fig6}
\end{figure}

\section{\bf Expansion in the Wannier representation}

In this section we are going to discuss the alternative perturbation expansion 
based on the decomposition (\ref{decomw}) of the three band Hamiltonian. 
The unperturbed Hamiltonian (\ref{h0}) consists of independent local
Hamiltonians $h_\bfl$ for each site which are easily diagonalized. In the one
hole sector the local Hamiltonian has two $S=1/2$ eigenstates where mutually
orthogonal linear combinations of $d$--hole and $w$--hole orbitals are 
occupied. The two hole sector contains a non--hybridized $S=1$ triplet and 
three hybridized singlets the lowest one of which is the Zhang--Rice singlet. 
In the three hole sector again two $S=1/2$ doublets are found. The four hole 
sector and the zero hole sector each contain one trivial $S=0$ state. The 
lower doublet in the one hole sector acts as the local ground state doublet 
of the undoped system. All the other states will show up as intermediate 
excited states at sufficiently high orders of the expansion with respect to 
the perturbation (\ref{hv}). We have diagonalized the local Hamiltonian 
numerically. A simple analytic formula cannot be obtained in the general case
since for the three singlets in the two hole sector a (3$\times$3)--matrix has
to be diagonalized. Simple analytic expressions for the solution in this sector
would be available only in the symmetric case $\dpd=U/2$. The perturbation 
expansion with respect to (\ref{hv}) was performed using a combination of 
symbolic and numerical routines. 

It is instructive to analyse the radius of convergence $t_{pd}^c$ of the 
local Hamiltonian $h_\bfl$ which does depend on $t_{pd}$ via (\ref{t0}). 
This radius of convergence can be determined from studying the branch points of
the eigenvalues of the local Hamiltonian in the complex $t_{pd}$--plane. 
Without going into any details we wish to summarize this analysis here by 
stating that $t_{pd}^c=0.469\,{\rm eV}$ for the values of $U$ and $\dpd$ given 
in (\ref{param}). This value agrees well with the values estimated from the 
Pad\'e approximants. This is not surprising since the expansion with respect to
the small perturbation (\ref{hv}) is expected to converge well and should not
much modify $t_{pd}^c$ as defined above from the local Hamiltonian.

\begin{figure}[h]
\begin{center}
\epsfxsize=12cm
\epsfbox{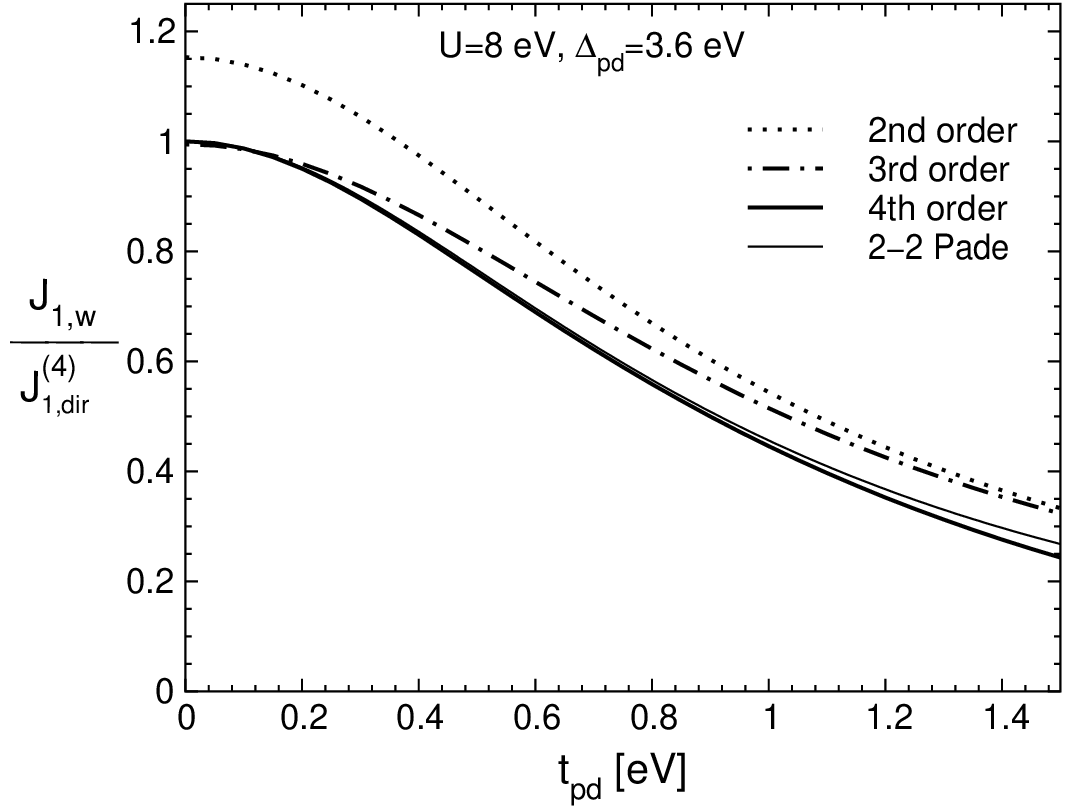}
\end{center}
\caption{Variation of $J_{1,w}/J_{1,{\rm dir}}^{(4)}$ with $t_{pd}$.}
\label{fig7}
\end{figure}
In what follows we will plot the variation of the various coupling constants
with the hopping $t_{pd}$ in analogy to the presentations in the previous
figures by measuring all couplings in units of their lowest order term in the
direct expansion (if not otherwise stated). 
Fig.~7 shows our results for the nearest neighbor exchange $J_{1,w}$. 
In the present context the leading contribution to $J_{1,w}$ is obtained 
from the simple second order hopping process described by the term 
$P_0V\!SV\!P_0$ of (\ref{perturb}). This second order contribution is depicted 
by the thick dotted line in Fig.~7. It is satisfying that this simple second 
order result reproduces quite nicely the decrease of 
$J_{1}/J_{1,{\rm dir}}^{(4)}$ with increasing $t_{pd}$ as given by the Pad\'e 
approximants of Fig.~1. On the other hand, there is, however, a systematic 
deviation in the overall size of the coupling; even for small $t_{pd}$ the 
coupling $J_{1,w}^{(2)}$ is too large by about $15\%$. The discrepancy 
at small $t_{pd}$ is largely reduced by taking into account the third order 
terms derived from $P_0V\!SV\!SV\!P_0$ and, finally, $J_{1,w}^{(4)}$ is 
in satisfying agreement with the 2-2~Pad\'e of the direct expansion. 
\begin{figure}[t]
\begin{center}
\epsfxsize=12cm
\epsfbox{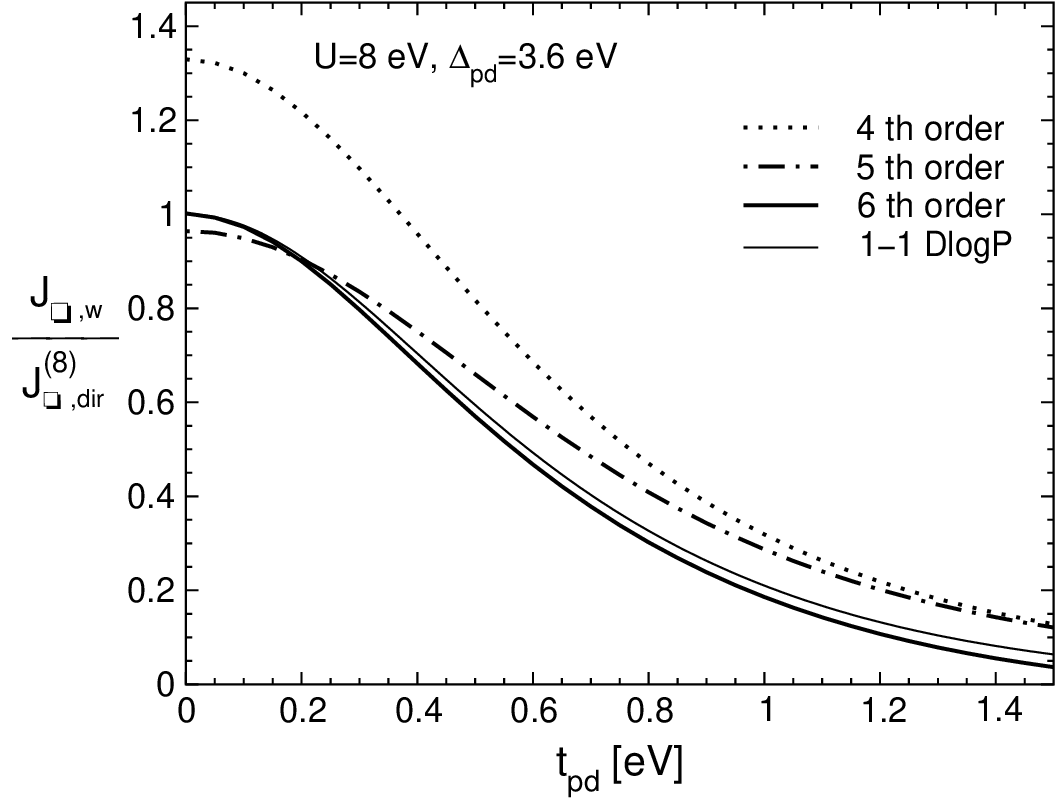}
\end{center}
\caption{Variation of $J_{\Box,w}/J_{\Box,{\rm dir}}^{(8)}$ with $t_{pd}$.}
\label{fig8}
\end{figure}
The deviation of $J_{1,w}^{(2)}$ from $J_{1,{\rm dir}}^{(4)}$ for small
$t_{pd}$ is explained quantitatively by re--expanding $J_{1,w}^{(2)}$ 
to second order with respect to $t_0$. Referring to (\ref{t0}) we obtain 
\begin{equation}\label{j1wasy}
J_{1,w}^{(2)}\sim(2T_{(0,0)}T_{(1,0)})^2J_{1,{\rm dir}}^{(4)}\qquad
(t_{pd}\to0)
\end{equation}
which explains the $15\%$ deviation because $(2T_{(0,0)}T_{(1,0)})^2=1.153$. 
Since $J_{1,w}^{(4)}$ collects all fourth order terms it has to coincide 
with $J_{1,{\rm dir}}^{(4)}$ after re--expanding it to $t_{pd}^4$. How this 
happens becomes particularly clear if one looks at the sum rule $s_{(1,0)}$ of
(\ref{sumrule}). This sum rule states that $2T_{(0,0)}T_{(1,0)}+r_{(1,0)}=-1$ 
if we denote by $r_{(1,0)}$ the sum of all terms in $s_{(1,0)}$ (infinitely
many) which don't contain $T_{(0,0)}$. Squaring this sum rule we obtain the
relation
\begin{equation}\label{sqsr}
(2T_{(0,0)}T_{(1,0)})^2+2(2T_{(0,0)}T_{(1,0)})r_{(1,0)}+r_{(1,0)}^2=1
\end{equation}
from which we can read off the contributions of various orders of the Wannier
expansion to $J_{1,{\rm dir}}^{(4)}$ in the limit of small $t_{pd}$. The first
term represents the contribution of $J_{1,w}^{(2)}$ discussed above.
The second term contains only one factor of $T_{(0,0)}$ and results from third
order terms in $V^w$ which due to $r_{(1,0)}=0.073775$ exhaust the 
relation (\ref{sqsr}) to $1-r_{(1,0)}^2=0.994557$; this explains why
$J_{1,w}^{(3)}$ is slightly smaller than $J_{1,{\rm dir}}^{(4)}$ in the
limit of small $t_{pd}$ (see Fig.~7). Finally, the term $r_{(1,0)}^2$ comes
from fourth order terms in $V^w$ which contribute only about $0.5\%$ for 
small $t_{pd}$ but change sign and get more important as $t_{pd}$ increases.

In our calculation of third and fourth order contributions to $J_{1,w}$
we have made extensive use of the sum rule $s_{(1,0)}$. In third order terms 
the exchange path for a spin flip process involves an arbitrary third copper 
ion site whose spin is not flipped. The sum of the spin flip amplitudes over 
all these third sites contains a lattice sum which is simply $r_{(1,0)}$. The
calculation of the fourth order is more involved since one has to discriminate
between spin flip processes which don't visit another site and those which
visit one or two more sites. The lattice sums appearing in this order cannot be
completely determined from sum rules, but sum rules considerably simplify 
their calculation. Four--spin terms which here appear in fourth order are
eliminated by averaging as described in the previous section.

Results for the four--spin coupling $J_{\Box,w}$ are shown in Fig.~8. 
The leading fourth order contribution again nicely reproduces qualitatively 
the decrease with increasing $t_{pd}$ known from Fig.~2. After the above 
discussion the deviation observed in the small $t_{pd}$ limit is not 
surprising. In fact, a quantitative understanding of this deviation follows 
from looking at the fourth power of the $s_{(1,0)}$ sum rule: 
$(2T_{(0,0)}T_{(1,0)}+r_{(1,0)})^4=1$. With $(2T_{(0,0)}T_{(1,0)})^4=1.329$
we understand why $J_{\Box,w}^{(4)}$ is about $33\%$ too large for small
$t_{pd}$. The substantial reduction of the deviation by the fifth order
contributions are also understood quantitatively from the identity
$(2T_{(0,0)}T_{(1,0)})^4+4(2T_{(0,0)}T_{(1,0)})^3r_{(1,0)}=0.964$ (see Fig.~8).
Including the sixth order terms we find the almost negligible deviation of
$(2T_{(0,0)}T_{(1,0)})^4+4(2T_{(0,0)}T_{(1,0)})^3r_{(1,0)}+
6(2T_{(0,0)}T_{(1,0)})^2r_{(1,0)}^2=1.0017$ and the overall agreement with 
the 1-1~DlogPad\'e is quite satisfying. 

\begin{figure}[h]
\begin{center}
\epsfxsize=11.5cm
\epsfbox{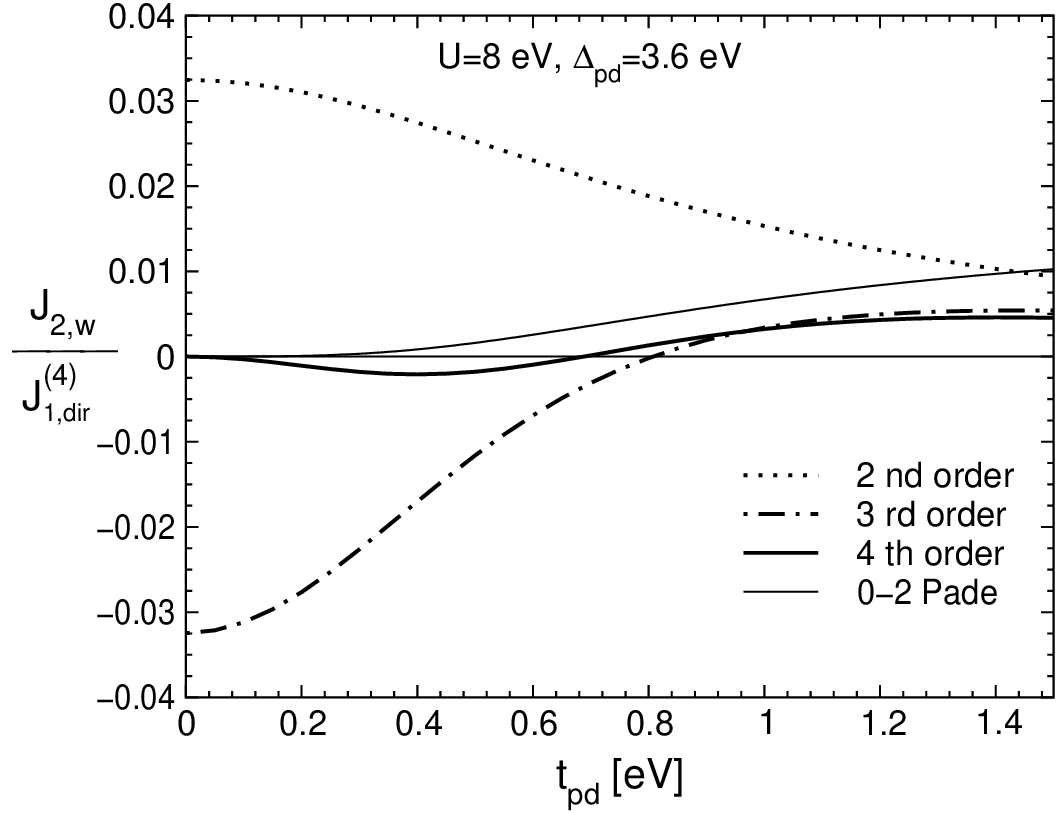}
\end{center}
\caption{Variation of $J_{2,w}/J_{1,{\rm dir}}^{(4)}$ with $t_{pd}$.}
\label{fig9}
\end{figure}
The analysis of the second and third neighbor exchange $J_2$ and $J_3$ is more
complicated in the Wannier representation since there are low order
contributions which have to be cancelled completely by higher order terms
before results of any significance emerge. We therefore show these couplings 
in Figs.~9 and 10 not in units of their eighth order counterparts from the 
direct expansion but in units of $J_{1,{\rm dir}}^{(4)}$. It is quite obvious
that for any lattice vector $\bfl$ there is a second order contribution 
$J_{\bfl,w}^{(2)}$ to the Heisenberg coupling between two spins separated
by $\bfl$ which in analogy to (\ref{j1wasy}) behaves like
\begin{equation}\label{jlwasy}
J_{\bfl,w}^{(2)}\sim(2T_{(0,0)}T_\bfl)^2J_{1,{\rm dir}}^{(4)}\qquad
(t_{pd}\to0).
\end{equation}
The cancellation of this contribution by higher order terms is understood by
invoking the sum rule $2T_{(0,0)}T_\bfl+r_\bfl=0$ for further neighbors which
squared gives the relation
\begin{equation}\label{sqsrl}
(2T_{(0,0)}T_\bfl)^2+2(2T_{(0,0)}T_\bfl)r_\bfl+r_\bfl^2=0.
\end{equation}

\begin{figure}[h]
\begin{center}
\epsfxsize=12cm
\epsfbox{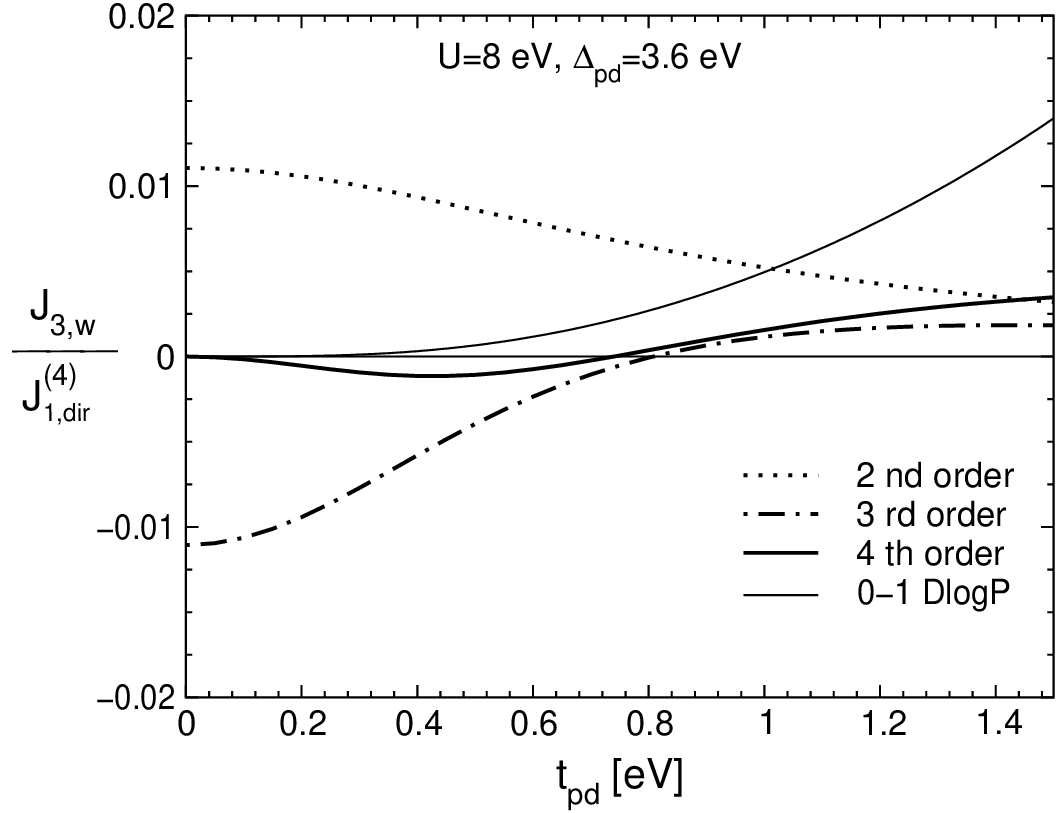}
\end{center}
\caption{Variation of $J_{3,w}/J_{1,{\rm dir}}^{(4)}$ with
$t_{pd}$.}
\label{fig10}
\end{figure}

The numbers $(2T_{(0,0)}T_{(1,1)})^2=0.0325$ and 
$(2T_{(0,0)}T_{(2,0)})^2=0.0111$ coincide perfectly with the behavior of
$J_{2,w}^{(2)}$ and $J_{3,w}^{(2)}$ for small $t_{pd}$ as shown in
Figs.~9 and 10. We also understand from (\ref{sqsrl}) why the inclusion of the
third order doesn't reduce the deviation from 0 but just changes its sign.
In fourth order Fig.~9 shows that in agreement with (\ref{sqsrl}) the
terms proportional to $t_{pd}^4$ in $J_{2,w}$ vanish. This is, however, 
only a partial solution of the cancellation problem since there are still 
terms proportional to $t_{pd}^6$ which according to Fig.~9 even have the wrong
sign and cancellation of which would only be achieved by extending the 
$V^w$ expansion to sixth order. For $t_{pd}>0.8\,{\rm eV}$ the third and
fourth order results shown in Figs.~9 and 10 at least have the right sign and 
the same order of magnitude as the Pad\'e estimates from the previous section. 
We have to conclude that the accurate determination of the further neighbor 
couplings $J_2$ and $J_3$ in the Wannier representation would be very 
demanding. This points at definite limitations of this approach. 

\section{\bf Conclusions}

In the present paper we have discussed the derivation of high precision 
effective spin Hamiltonians for the low energy sector of a three band model 
for \cuotwo--planes. By two methods we have demonstrated that it is possible 
to overcome the convergence problems of the $t_{pd}$ perturbation series. Using
the direct expansion with respect to $t_{pd}$ we have derived precise values
for the most important couplings via Pad\'e approximants. The direct expansion 
has the advantage of a particularly simple unperturbed Hamiltonian and a very 
nicely localized perturbative Hamiltonian which makes high order symbolic 
expansions feasible. Using the Wannier representation we have confirmed the
results from the direct expansion by a method with much better convergence
properties. The expansion in the Wannier representation is, however, rendered
more difficult by a more complicated unperturbed Hamiltonian and a less well
localized perturbative Hamiltonian and by the necessity of a non--symbolic 
(i.e. numerical) series expansion. We have also shown that for precise values 
of the coupling constants the leading orders of the Wannier expansion are not 
sufficient.

The work in the present paper was confined to the most rudimentary three band
model since our main goal was to demonstrate the feasibility of the derivation
of accurate effective Hamiltonians. Nevertheless we will use our results here 
for a fit of the couplings $J_1=151.9\,{\rm meV}$ and $J_\Box/J_1=0.24$ 
extracted recently from a fit to the experimental dispersion of $La_2CuO_4$ 
\cite{coldea} using self consistent spin--wave theory \cite{katanin}. 
Assuming the typical (though somewhat arbitrary) model parameters 
$U=8\,{\rm eV}$ and $\dpd=3.6\,{\rm eV}$ from (\ref{param}) we obtain from 
the value $J_1=151.9\,{\rm meV}$ the estimate 
$1.422\,{\rm eV}\le t_{pd}\le1.454\,{\rm eV}$ for the hopping parameter of 
the minimum model showing an uncertainty in $t_{pd}$ of $2\%$ due to the 
uncertainty of our Pad\'e extrapolations. With $t_{pd}$ in this range our 
estimate for $J_\Box$ results in $0.19 \le J_\Box/J_1 \le 0.25$ which is in 
good agreement with the result from \cite{katanin}.

For proper applications to cuprate materials this work will have to be 
extended to more realistic three band models including, in particular, a 
direct oxygen--oxygen hopping $t_{pp}$ \cite{mcmahan,hybertsen}. The relevance 
of four--spin exchange has been stressed also for the two--leg ladder system
$La_6Ca_8Cu_{24}O_{41}$ \cite{matsuda} to which the analysis presented here
can be applied as well. 

\section*{Acknowledgement}
The authors gratefully acknowledge useful discussions with Christian Knetter 
and Kai Schmidt. This work was performed within a research project supported
by the German--Israeli Foundation.
\newpage

\begin{appendix}
\renewcommand\thesection{Appendix \Alph{section}} 
\renewcommand\theequation{\Alph{section}\@.{\arabic{equation}}} 
\makeatletter 
\@addtoreset{equation}{section} 
\makeatother
\section{}

This Appendix contains the more voluminous formulae from the direct expansion 
of section IV. These formulae can be easily used to derive the Pad\'e 
approximants discussed in section IV.

With (\ref{jdir}) the Taylor series for the nearest neighbor exchange coupling 
is 
\begin{eqnarray}\label{J1Ordn10}
J_{1,{\rm dir}}^{\vphantom{4}}
&=&J_{1,{\rm dir}}^{(4)}\Big[\,1-t_{pd}^2\,
\frac{4\,\left( 5\,U + 2\,\dpd \right) }
     {\dpd^2\,\left( U + \dpd \right)}+\nonumber\\
&&\,t_{pd}^4\,
\frac{801\,U^3 + 164\,U^2\,\dpd - 24\,U\,\dpd^2 - 12\,\dpd^3}
     {2\,U^2\,\dpd^4\,\left( U + \dpd \right) }-\nonumber\\
&&\,t_{pd}^6\,
\frac{8505\,U^4 + 9602\,U^3\,\dpd + 908\,U^2\,\dpd^2 - 240\,U\,\dpd^3 
- 48\,\dpd^4}{U^2\,\dpd^6\,{\left( U + \dpd \right) }^2}+\nonumber\\
&&\,t_{pd}^8(758199\,U^7 + 1587453\,U^6\,\dpd + 890808\,U^5\,\dpd^2 
+ 52603\,U^4\,\dpd^3 \nonumber\\
&&- 6611\,U^3\,\dpd^4 + 4566\,U^2\,\dpd^5 + 2559\,U\,\dpd^6 + 483\,\dpd^7)
/\nonumber\\
&&(4\,U^4\,\dpd^8\,{\left( U + \dpd \right) }^3)
+O(t_{pd}^{10})\,\Big].
\end{eqnarray}
The series for the four--spin coupling with the leading contribution
(\ref{jsquare}) is given by
\begin{eqnarray}\label{JsqOrdn12}
J_{\Box,{\rm dir}}^{\vphantom{4}}
&=&J_{\Box,{\rm dir}}^{(8)}\Big[\,1-t_{pd}^2\frac
{4\,\left( 11\,U^3 + 14\,U^2\,\dpd + 8\,U\,\dpd^2 + 2\,\dpd^3 \right)}
{\dpd^2\,\left( U + \dpd \right) \,\left( U^2 + U\,\dpd + 
\dpd^2 \right) }+\nonumber\\
&&t_{pd}^4(56569\,U^7 + 161892\,U^6\,\dpd + 168480\,U^5\,\dpd^2 
           + 76092\,U^4\,\dpd^3 \nonumber\\
&&         + 9096\,U^3\,\dpd^4 - 7008\,U^2\,\dpd^5 - 3960\,U\,\dpd^6 
           - 792\,\dpd^7)/\nonumber\\
&&(40\,U^2\,\dpd^4\,( U + \dpd)^3\,( U^2 + U\,\dpd +\dpd^2))\nonumber\\
&&-t_{pd}^6 
(410565\,U^8 + 1487797\,U^7\,\dpd + 2034672\,U^6\,\dpd^2 \nonumber\\
&&+ 1264452\,U^5\,\dpd^3 + 296152\,U^4\,\dpd^4 - 48240\,U^3\,\dpd^5 
\nonumber\\
&&- 49264\,U^2\,\dpd^6 - 13464\,U\,\dpd^7 - 1584\,\dpd^8)/\nonumber\\
&&(10\,U^2\,\dpd^6\,( U + \dpd)^4\,( U^2 + U\,\dpd + \dpd^2 ))
+O(t_{pd}^8)\,\Big].
\end{eqnarray}
The leading contribution to the four--spin coupling (\ref{timesex}) is
\begin{eqnarray}\label{JxOrdn12}
\!\!\!\!\!\!\!\!\!\!J_\times^{(12)}
&=&2\,t_{pd}^{12}\,\big(\,
( 489\,U^7 + 1016\,U^6\,\dpd - 72\,U^5\,\dpd^2 - 
2232\,U^4\,\dpd^3 \nonumber\\
&&  - 3392\,U^3\,\dpd^4 - 2784\,U^2\,\dpd^5 - 1280\,U\,\dpd^6 - 
256\,\dpd^7 ) \big)/\nonumber\\
&&\big(U^5\,\dpd^{11}\,( U + \dpd)^2\big)
\end{eqnarray}
With (\ref{jtwo}) and (\ref{jthree}) the series for the second and third
neighbor two--spin couplings are given by
\begin{eqnarray}\label{J2Ordn12}
J_{2,{\rm dir}}^{\vphantom{4}}
&=&J_{2,{\rm dir}}^{(8)}\Big[\,1-\nonumber\\
&&t_{pd}^2\frac{4\,( 142\,U^4 + 169\,U^3\,\dpd + 36\,U^2\,\dpd^2 + 
10\,U\,\dpd^3 + 2\,\dpd^4 ) }{\dpd^2\,( U + \dpd) \,
    ( 11\,U^3 + 4\,U^2\,\dpd + 2\,U\,\dpd^2 + \dpd^3) }+\nonumber\\
&&t_{pd}^4\big(82083\,U^7 + 171784\,U^6\,\dpd + 99154\,U^5\,\dpd^2 + 
10848\,U^4\,\dpd^3 \nonumber \\
&&+ 1420\,U^3\,\dpd^4 + 290\,U^2\,\dpd^5 + 
    120\,U\,\dpd^6 + 24\,\dpd^7\big)/\nonumber\\
&&\big(4\,U^2\,\dpd^4\,( U + \dpd)^2\,
( 11\,U^3 + 4\,U^2\,\dpd + 2\,U\,\dpd^2 + \dpd^3 ) \big)\nonumber\\
&&+O(t_{pd}^6)\,\Big]
\end{eqnarray}
and 
\begin{eqnarray}\label{J3Ordn12}
J_{3,{\rm dir}}^{\vphantom{4}}
&=&J_{3,{\rm dir}}^{(8)}\Big[\,1-\nonumber\\
&&t_{pd}^2\frac{2\,( 72\,U^4 + 94\,U^3\,\dpd + 39\,U^2\,\dpd^2 + 
20\,U\,\dpd^3 + 4\,\dpd^4) }{\dpd^2\,( U + \dpd) \,
( 3\,U^3 + 2\,U^2\,\dpd + 2\,U\,\dpd^2 + \dpd^3) }+\nonumber\\
&&t_{pd}^4\big(47947\,U^7 + 111156\,U^6\,\dpd + 90704\,U^5\,\dpd^2 + 
43130\,U^4\,\dpd^3 \nonumber\\
&&+ 21424\,U^3\,\dpd^4 + 8174\,U^2\,\dpd^5 + 2632\,U\,\dpd^6 + 
440\,\dpd^7\big)/\nonumber\\
&&\big(8\,U^2\,\dpd^4\,( U + \dpd)^2\,
( 3\,U^3 + 2\,U^2\,\dpd + 2\,U\,\dpd^2 + \dpd^3) \big)\nonumber\\
&&+O(t_{pd}^6)\,\Big].
\end{eqnarray}
The 5 six--spin invariants for a hexagonal plaquette are
\begin{eqnarray}\label{opsix}
O_1&=&(\bfS_1\cdot\bfS_2)(\bfS_3\cdot\bfS_4)(\bfS_5\cdot\bfS_6)+
(\bfS_2\cdot\bfS_3)(\bfS_4\cdot\bfS_5)(\bfS_6\cdot\bfS_1)\nonumber\\
O_2&=&(\bfS_1\cdot\bfS_4)(\bfS_2\cdot\bfS_6)(\bfS_3\cdot\bfS_5)+
(\bfS_2\cdot\bfS_5)(\bfS_3\cdot\bfS_1)(\bfS_4\cdot\bfS_6)\nonumber\\
&&+(\bfS_3\cdot\bfS_6)(\bfS_4\cdot\bfS_2)(\bfS_5\cdot\bfS_1)\nonumber\\
O_3&=&(\bfS_1\cdot\bfS_4)(\bfS_2\cdot\bfS_5)(\bfS_3\cdot\bfS_6)\nonumber\\
O_4&=&(\bfS_1\cdot\bfS_2)(\bfS_3\cdot\bfS_6)(\bfS_4\cdot\bfS_5)+
(\bfS_2\cdot\bfS_3)(\bfS_4\cdot\bfS_1)(\bfS_5\cdot\bfS_6)\nonumber\\
&&+(\bfS_3\cdot\bfS_4)(\bfS_5\cdot\bfS_2)(\bfS_6\cdot\bfS_1)\nonumber\\
O_5&=&(\bfS_1\cdot\bfS_2)(\bfS_3\cdot\bfS_5)(\bfS_4\cdot\bfS_6)+
(\bfS_2\cdot\bfS_3)(\bfS_4\cdot\bfS_6)(\bfS_5\cdot\bfS_1)\nonumber\\
&&+(\bfS_3\cdot\bfS_4)(\bfS_5\cdot\bfS_1)(\bfS_6\cdot\bfS_2)+
(\bfS_4\cdot\bfS_5)(\bfS_6\cdot\bfS_2)(\bfS_1\cdot\bfS_3)\nonumber\\
&&+(\bfS_5\cdot\bfS_6)(\bfS_1\cdot\bfS_3)(\bfS_2\cdot\bfS_4)+
(\bfS_6\cdot\bfS_1)(\bfS_2\cdot\bfS_4)(\bfS_3\cdot\bfS_5).
\end{eqnarray}
The leading contribution to the six--spin coupling in Eq.~(\ref{sixex}) is
found to be 
\begin{equation}\label{Jsix}
J_{\sqsubset\!\sqsupset}^{(12)}=
\frac{336\,t_{pd}^{12}\,( U + \dpd) \,( 3\,U^4 + 6\,U^3\,\dpd + 
8\,U^2\,\dpd^2 + 6\,U\,\dpd^3 + 3\,\dpd^4)}{U^5\,\dpd^{11}}.
\end{equation}

\end{appendix}


\begin{thebibliography}{99999999}

\bibitem{bednorz} {J.G. Bednorz and K.A. M\"uller, Z. Phys. B {\bf 64}, 189
(1986).}
\bibitem{emery} {V.J. Emery, Phys. Rev. Lett. {\bf 58}, 2794 (1987).}
\bibitem{andersen} {Band structure calculations imply that a quantitative
description of the electronic structure requires as many as eight bands, see
e.g.: O.K. Andersen et al., J. Phys. Chem. Solids {\bf 56}, 1573 (1995).}
\bibitem{zhang1} {F.C. Zhang and T.M. Rice, Phys. Rev. B {\bf 37}, 
3759 (1988).}
\bibitem{anderson} {P.W. Anderson, Science {\bf 235}, 1196 (1987).}
\bibitem{takahashi} {For a nice presentation of the method, see: M. Takahashi,
J. Phys. C: Solid State Phys. {\bf 10}, 1289 (1977).}
\bibitem{stein} {J. Stein, O. Entin--Wohlman and A. Aharony, Phys. Rev. B 
{\bf 53}, 775 (1996); J. Stein, Phys. Rev. B {\bf 53}, 785 (1996).}
\bibitem{yildirim} {T. Yildirim et al., Phys. Rev. Lett. {\bf 73}, 2919 (1994);
T. Yildirim et al., Phys. Rev. B {\bf 52}, 10239 (1995); O. Entin--Wohlman,
A.B. Harris and A. Aharony, Phys. Rev. B {\bf 53}, 11661 (1996).}
\bibitem{schmidt} {see also: H.J. Schmidt and Y. Kuramoto, Physica B {\bf 163},
443 (1990), and more recently: Y. Mizuno et. al., J. Low Temp. Ph. {\bf 117}, 
389 (1999); analogous results for the single band Hubbard model are obtained
in \cite{macdonald}.}
\bibitem{katanin} {A.A. Katanin and A.P. Kampf, cond-mat/0111533.}
\bibitem{coldea} {R. Coldea et al., Phys. Rev. Lett. {\bf 86}, 5377 (2001).}
\bibitem{dagotto} {For a review on the modeling of $CuO_2$--planes, see: 
E. Dagotto, Rev. Mod. Phys. {\bf 66}, 763 (1994).}
\bibitem{mcmahan} {A.K. McMahan, R.M. Martin and S. Satpathy, Phys. Rev. B
{\bf 38}, 6650 (1988).}
\bibitem{hybertsen} {M.S. Hybertsen, M. Schl\"uter and N.E. Christensen,
Phys. Rev. B {\bf 39}, 9028 (1989).}
\bibitem{reiter1} {V.J. Emery and G. Reiter, Phys. Rev. B {\bf 38}, 
4547 (1988).}
\bibitem{reiter2} {V.J. Emery and G. Reiter, Phys. Rev. B {\bf 38}, 
11938 (1988).}
\bibitem{zhang2} {F.C. Zhang and T.M. Rice, Phys. Rev. B {\bf 41}, 
7243 (1990).}
\bibitem{reiter3} {V.J. Emery and G. Reiter, Phys. Rev. B {\bf 41}, 
7247 (1990).}
\bibitem{pang} {H.B. Pang, T. Xiang, Z.B. Su and L. Yu, Phys. Rev. B {\bf 41},
7209 (1990).}
\bibitem{zaanen} {J. Zaanen and A.M. Ole\'s, Phys. Rev. B {\bf 37}, 9423
(1988).}
\bibitem{lovtsov} {S.V. Lovtsov and V.Y. Yushankhai, Physica C {\bf 179},
159 (1991).}
\bibitem{schuettler} {H.B. Sch\"uttler and A.J. Fedro, Phys. Rev. B {\bf 45}, 
7588 (1992)}
\bibitem{jefferson} {J.H. Jefferson et al., Phys. Rev. B {\bf 45}, 7959
(1992).}
\bibitem{hayn} {R. Hayn, V. Yushankhai and S. Lovtsov, Phys. Rev. B {\bf 47},
5253 (1993).}
\bibitem{belinicher} {V.I. Belinicher and A.L. Chernyshev, Phys. Rev. B 
{\bf 49}, 9746 (1994).}
\bibitem{feiner} {L.F. Feiner et al., Phys. Rev. B {\bf 53}, 8751 (1996).}
\bibitem{oles} {A.M. Ole\'s, Phys. Rev. B {\bf 41}, 2562 (1990).}
\bibitem{macdonald} {A.H. MacDonald, S.M. Girvin and D. Yoshioka, 
Phys. Rev. B {\bf 41}, 2565 (1990).}
\bibitem{spat} {Products of two triple products like 
$[(\bfS_1\times\bfS_2)\cdot\bfS_3][(\bfS_4\times\bfS_5)\cdot\bfS_6]$ don't
represent additional invariants since they can be expressed by scalar 
products.}
\bibitem{matsuda} {M. Matsuda et al., J. Appl. Phys. {\bf 87}, 6271 (2000);
Phys. Rev. B {\bf 62}, 8903(2000).}

\end{thebibliography}
\end{document}